\documentclass[apjl]{emulateapj}
\usepackage{natbib}
\usepackage{epsfig}
\usepackage{apjfonts}
\usepackage{subfigure}
\bibpunct{(}{)}{;}{a}{}{,}

\shorttitle{New Taurus VLM Binaries}
\shortauthors{Konopacky et al.}
\begin{document}

\title{New Very Low Mass Binaries in the Taurus Star-Forming Region}
\author{Q.M. Konopacky\altaffilmark{1,2},
  A.M. Ghez\altaffilmark{1,2}, E.L. Rice\altaffilmark{1}, and
   G. Duch\^{e}ne\altaffilmark{3}}
\altaffiltext{1}{UCLA Division of Astronomy and Astrophysics, Los Angeles, CA 90095-1562; quinn, ghez@astro.ucla.edu}
\altaffiltext{2}{Institute of Geophysics and Planetary Physics, University of
California, Los Angeles, CA 90095-1565}
\altaffiltext{3}{Laboratoire d'Astrophysique,
Observatoire de Grenoble, Universite Joseph Fourier - BP 53,
F-38041, Grenoble Cedex 9, France; Gaspard.Duchene@obs.ujf-grenoble.fr}
\begin{abstract}
    We surveyed thirteen very low mass (VLM; M $\lesssim$0.2
    M$_{\sun}$) objects in the Taurus star-forming region
    using near-infrared diffraction-limited imaging techniques
    on the W.M. Keck I 10 m telescope.  Of these thirteen,
    five were found to be binary, with separations ranging
    from 0$\farcs$04 to 0$\farcs$6 and flux ratios from 1.4 to
    3.7.  In all cases, the companions are likely to be
    physically associated with the primaries (probability $\gtrsim$4$\sigma$).
    Using the theoretical models of Baraffe et al. (1998), we
    find that all five new companions, as well as one of the
    primaries, are likely brown dwarfs.  The discovery of
    these systems therefore increases the total number of
    known, young VLM binaries by $\thicksim$50$\%$.  These new
    systems, along with other young VLM binaries from the
    literature, have properties that differ significantly from
    older field VLM binaries in that the young systems have
    wider separations and lower mass ratios, supporting the
    idea that VLM binaries undergo significant dynamical
    evolution $\thicksim$5 - 10 Myr after their formation.
    The range of separations of these binaries, four of which
    are over 30 AU, argues against the ejection scenario of
    brown dwarf formation.  While several of the young, VLM
    binaries discovered in this study have lower binding
    energies than the previously suggested minimum for VLM
    binaries, the apparent minimum is still significantly
    higher than that found among higher mass binaries.  We
    suggest that this discrepancy may be due to the small mass
    of a VLM binary relative to the average perturbing star,
    leading to more substantial changes in their
    binding energy over time.

\end{abstract}


\section{Introduction}

   Since the first observational detection of brown dwarfs -
 objects whose mass ($\lesssim$0.08 M$_{\sun}$) is too small
 for them to achieve hydrogen fusion but larger than the
 majority of known planets - the mechanism for their formation
 has been a hotly pursued and elusive puzzle.  Current
 theories include formation of very low-mass cores in
 turbulent clouds (e.g. Padoan and Nordlund, 2004),
 additional fragmentation in higher-mass cores (e.g. Boss
 2002), fragmentation out of high mass discs (e.g. Rice et
 al. 2003), premature ejection from a natal cloud core
 (e.g. Bate et al. 2002), and photo-erosion of cores by nearby
 OB stars (e.g. Whitworth and Zinnecker, 2004).  Over the
 past few years, extensive surveys of star-forming regions
 have been undertaken to identify larger numbers of VLM
 members (e.g. Zapatero Osorio et al. 2000, Ardila et
 al. 2000, Luhman 2004a, Guieu et al. 2006).  By studying the
 properties of brown dwarfs and VLM stars at very young ages,
 one may gain substantial insight into their origins.
 
One such property that is useful for gaining this insight is
multiplicity fractions of VLM stars and brown dwarfs.  Many
formation scenarios make predictions as to the percentage of
these objects that would be expected to be found in binaries.
A number of field VLM star and brown dwarf multiplicity
studies have been carried out and have found very low overall
binary fractions ($\thicksim$5-10$\%$), mass ratio
distributions that are strongly peaked toward equal masses,
and a sharp decline in the binary fraction beyond separations
of 20 AU (e.g. Burgasser et al., 2003, Close et al., 2003,
Bouy et al., 2003, Gizis et al. 2003).  When compared with
higher mass studies of field stars (e.g. Duquennoy $\&$ Mayor
1991), these suggest that the maximum binary separation
decreases with mass.  However, these results may be affected
by dynamical evolution given the substantial age of the
targets.  Indeed, multiplicity studies of stellar objects have
shown evidence for dynamical evolution, as the binary fraction
of these objects is generally found to be much higher in young
star-forming regions than in the field (e.g. Ghez et
al. 1993, Leinert et al. 1993, Simon et al. 1995).  Thus,
surveys of younger ($\thicksim$1 - 5 Myr) VLM systems are much
more likely to reveal the pristine outcome of the brown dwarf
formation process.  Such work has just begun over the last few
years and includes studies in Upper Scorpius (distance = 145
pc, age $\thicksim$ 5 Myr; Kraus et al. 2005, Bouy et
al. 2006) and Taurus (distance = 140 pc, age $\thicksim$1-5
Myr; Kraus et al. 2006).  To date, only a handful of young,
VLM binaries have been detected.
 
In this paper, we report the discovery of five new VLM
binaries in Taurus.  In contrast to field VLM binaries, the
majority of these new binaries have separations greater than
20 AU.  These new binaries increase the number of known young
systems by $\thicksim$50$\%$ and suggest a higher binary
fraction among young, VLM objects.  In $\S$2, we describe our
observations and analysis, in $\S$3 we present our
results, in $\S$4 we compare our results to other work and
discuss the implications for brown dwarf
formation scenarios, and in $\S$5 we summarize our findings.
    
\section{Observations and Data Analysis}\label{obs}
   
\begin{deluxetable*}{lccccc}
\tabletypesize{\scriptsize}
\tablecolumns{6}
\tablewidth{0pc}
\tablecaption{Observation Summary}
\tablehead{
\colhead{Target} &
\colhead{Sp. Type} & \colhead{K (mag)\tablenotemark{a}} & \colhead{Date of
  Observation} & \colhead{Calibrator} & \colhead{Ref\tablenotemark{b}}
}
\startdata
CFHT-Tau 7 & M6.5 & 10.4 & 2005 Nov 13 & FZ Tau & 2 \\
CHFT-Tau 17 & M5.75 & 10.8 & 2005 Nov 13 & DO Tau & 2 \\
CFHT-Tau 18 &  M6.0 & 8.7 & 2005 Nov 13 & FZ Tau & 2 \\
CFHT-Tau 19 & M5.25 & 10.5 & 2005 Nov 13 & SAO 76547 & 2 \\
CFHT-Tau 20 &  M5.5 & 9.8 & 2005 Nov 12 & FZ Tau & 2 \\
CFHT-Tau 21 & M1.25 & 9.0 & 2005 Nov 13 & SAO 76547 & 2 \\
J04161210 &  M4.75 & 10.3 & 2004 Dec 19 & CW Tau & 1 \\
J04213459 &  M5.5 & 10.4 & 2005 Nov 13 & SAO 76547 & 1 \\
J04284263 &  M5.25 & 10.5 & 2004 Dec 19 & SAO 76628 & 1 \\
J04380083 &  M7.25& 10.1 & 2004 Dec 19 & DO Tau & 1 \\
J04403979 &  M5.5 & 10.2 & 2005 Nov 13 & DO Tau & 1 \\
J04442713 &  M7.25 & 10.8 & 2005 Nov 13 & SAO 76727 & 1 \\
J04554535 &  M4.75 & 10.5 & 2005 Nov 12 & SU Aur & 1 \\
\enddata
\tablenotetext{a}{From the 2MASS point source catalog}
\tablenotetext{b}{Source identified as VLM object by Luhman
  (2004; Ref 1) or Guieu et al. (2006; Ref 2)}
\label{obssum}
\end{deluxetable*}

   The observations were made using the Keck I 10 m telescope
 with the facility Near Infrared Camera (NIRC, Matthews $\&$
 Soifer 1994, Matthews et al. 1996) in speckle imaging mode on
 2004 December 19 and 2005 November 12-13.  In its high
 angular resolution modes, NIRC has a pixel scale of 20.45
 $\pm$ 0.03 mas/pixel.  The total field of view of NIRC in
 this mode is 5$\farcs$2.  For each target, four to six stacks
 of 190 images, each 0.137 seconds integration time, were obtained through
 the $K$ band-pass filter ($\lambda_o$ = 2.2 $\mu$m,
 $\Delta\lambda$ = 0.4 $\mu$m).  Along with these target
 stacks, we obtained stacks of darks, sky, and point source
 calibrators, all of which are used in the reduction process.
 Using these stacks, we generate power spectra of each of the
 targets using Fourier transform techniques (Labeyrie 1970).
 The procedure of speckle data analysis, including the
 creation of power spectra, are described in some detail in
 Konopacky et al. (2007) and Ghez et al. (1995), and this
 study uses the same approach.  Binary star power spectra
 exhibit a characteristic sinusoidal pattern, which is used to
 obtain the binary separation, position angle, and flux ratio.
 Shift-and-Add images (Christou 1991) were also generated from
 each data set to enhance our sensitivity to high flux ratio,
 wide binary systems.  Table \ref{obssum} lists all targets
 observed, as well as the point sources used for calibration.

 All targets were selected from Luhman (2004a) and Guieu et
 al. (2006) based upon their $K$ band magnitudes.  Theses two
 studies found 22 and 17 new VLM objects in the Taurus
 star-forming region, respectively.  Among the stars from
 these studies, fifteen are brighter than the speckle
 magnitude limit of K $\lesssim$11, which is required to
 achieve sufficient signal-to-noise in a single short exposure
 image.  During three nights at Keck, we observed thirteen of
 these targets, six from Guieu et al. (2006) and seven from
 Luhman (2004a) (J04554757 and J04555288 from Luhman (2004a)
 were not observed).  While there has been a range of
 definitions of VLM objects in the literature, in this study
 we define a VLM object as an object with a mass $\lesssim$0.2
 M$_{\sun}$.  One of the targets observed during this program,
 CFHT-Tau 21, has a spectral type of M1.25 and is not a VLM
 object by our definition.  Thus, we report it here for
 completeness, but exclude it from further analysis.  Table 1
 lists the spectral type and total $K$ band magnitude for each
 of the target stars.  Our VLM sample has spectral types that
 range from M4.75 to M7.25 and corresponding masses that range
 from 0.2 M$_{\sun}$ to 0.05 M$_{\sun}$ at the average age of
 Taurus ($\thicksim$3 Myr).

\section{Results}
\begin{deluxetable*}{lcccccccc}
\tabletypesize{\scriptsize}
\tablecolumns{9}
\tablewidth{0pc}
\tablecaption{Binary System Properties}
\tablehead{
\colhead{Target} &
\colhead{Sep. (``)} & \colhead{P.A.} & \colhead{Flux} &
\colhead{M$_{K}$} & \colhead{M$_{K}$} &
\colhead{Prob. of} & \colhead{$\thicksim$Primary} & \colhead{$\thicksim$Secondary}\\
\colhead{} &\colhead{} & \colhead{(degrees)} & \colhead{Ratio
  (K)} & \colhead{Primary} & \colhead{Secondary}  & \colhead{Background Object} & \colhead{Mass
  (M$_\odot$)} & \colhead{Mass (M$_\odot$)} 
}
\startdata
CFHT-Tau 7 & 0.224 $\pm$ 0.002 & 292.92 $\pm$ 0.17 & 1.40
$\pm$ 0.09 & 5.26 $\pm$ 0.18 & 5.61 $\pm$ 0.20 & 4.6 x 10$^{-6}$ & 0.07 & 0.06 \\
CHFT-Tau 17 & 0.575 $\pm$ 0.002 & 235.37 $\pm$ 0.31 & 3.70
$\pm$ 0.55 & 4.62 $\pm$ 0.18 & 6.08 $\pm$ 0.27 & 1.1 x 10$^{-4}$ & 0.1 & 0.06 \\
CFHT-Tau 18 & 0.216 $\pm$ 0.002 & 268.56 $\pm$ 0.34 & 2.28
$\pm$ 0.25 & 2.92 $\pm$ 0.18 & 3.77 $\pm$ 0.19 & 1.7 x 10$^{-6}$ & 0.1 & 0.06 \\
J04284263 & 0.621 $\pm$ 0.007 & 349.97 $\pm$ 0.83 & 2.29 $\pm$
0.39 & 5.07 $\pm$ 0.18 & 5.97 $\pm$ 0.26 & 6.3 x 10$^{-5}$ & 0.15 & 0.06 \\
J04403979 & 0.041 $\pm$ 0.003 & 289.98 $\pm$ 4.59 & 2.08 $\pm$
0.29 & 4.59 $\pm$ 0.29 & 5.68 $\pm$ 0.94 & 3.3 x 10$^{-7}$ & 0.15 & 0.08 \\ 
\enddata
\end{deluxetable*}

     Of the thirteen targets observed, five were found to be
binaries.  The parameters of each binary system are summarized
in Table 2.  As listed in Table 3, sensitivity estimates show
that, in general, companions with $\Delta$K = 3 could have
been detected at the 3$\sigma$ confidence level, all the way
down to an angular separation of 0$\farcs$02, the minimum
detectable separation using the speckle technique (see Ghez et
al. 1993, Konopacky et al. 2007).  Using the models of Baraffe
et al. (1998, $\alpha$ = 1.0), these 3$\sigma$ magnitude
difference sensitivities can be converted to an estimate of
detectable q (where q $\equiv$ M$_{secondary}$ /
M$_{primary}$).  For all of the targets in the sample, the
$\Delta$K to q conversion is quite similar and is
well-described by the equation q = 0.077$\Delta$K$^{2}$ -
0.526$\Delta$K + 0.966.  This survey therefore was generally
sensitive to q $\thicksim$0.23 at $\thicksim$3 AU, a region of
parameter space completely inaccessible to many past surveys.
Figure 1 shows the 3$\sigma$ sensitivity limits on q versus
distance from the primary source.

\begin{figure}
\epsscale{1.0}
\plotone{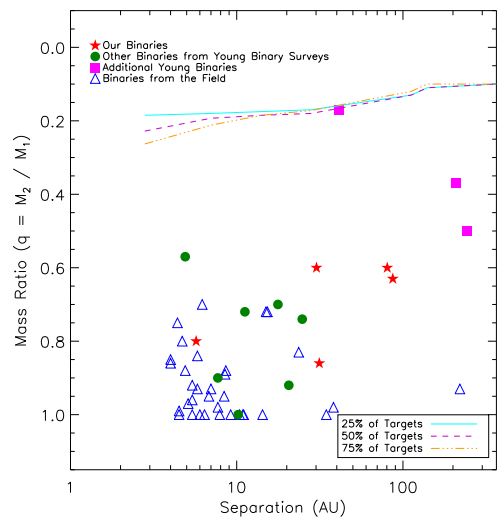}
\caption{The mass ratio q versus the separation (in AU) of the
  VLM binaries found in our survey and those found in other
  studies of both young and old VLM binaries.  This figure
  demonstrates the differences in the parameters of the old
  and the young samples.  The truncation of
  field VLM binaries at separation $>$ 20 AU and q $<$ 0.8 is not
  seen amongst young VLM binaries.  Also shown are the 3$\sigma$
  sensitivity limits of our survey.  The lines represent the
  sensitivities of the best 25, 50, and 75$\%$ of our sample.
  Quoted sensitivities are from the 50$\%$ curve.}
\label{young_compare}
\end{figure}

Because only one epoch of data on these targets has thus far
been obtained, the physical association of the five binaries
via common proper motions cannot be confirmed.  Instead, Table
2 lists the probability of association based upon the number
of sources of comparable brightness to the secondaries in the
region and the separation of the two stars.  These
probabilities were calculated following the method of Brandner
et al. (2000), using their equation 1.  The 2MASS All-Sky
Point Source Catalog Statistics
Service\footnote{http://irsa.ipac.caltech.edu/applications/Stats/}
was used to find the number of sources of comparable
brightness to the secondary within one square degree centered
on the primary.  These source densities were then used to
calculate the probability that these objects are chance
background sources.  These probabilities are given in Table
2. We find that all are likely to be physically associated
with probabilities of the secondary being a background object
of $\lesssim$ 10$^{-4}$, and thus conclude that the detected
binaries are physically associated with each other.

\begin{deluxetable}{lccccccc}
\tabletypesize{\scriptsize}
\tablecolumns{8}
\tablewidth{0pc}
\tablecaption{Limits on Mass Ratio for Undetected Companions to Single Stars}
\tablehead{
\colhead{Target} &
\colhead{Est.} & \colhead{0$\farcs$02} &
\colhead{0$\farcs$05} & \colhead{0$\farcs$1} &
\colhead{0$\farcs$2} & \colhead{0$\farcs$8} & \colhead{1$\farcs$0}\\
\colhead{ } &
\colhead{Mass (M$_{\sun}$)} & \colhead{ } &
\colhead{ } & \colhead{ } &
\colhead{ } & \colhead{ } & \colhead{ }\\
}
\startdata
CFHT-Tau 19 & 0.15 & 0.28 & 0.15 & 0.13 & 0.12 & 0.10 & 0.10  \\
CFHT-Tau 20 & 0.12 & 0.32 & 0.32 & 0.120 & 0.17 & 0.10 & 0.10 \\
CFHT-Tau 21 & 0.70 & 0.79 & 0.35 & 0.24 & 0.19 & 0.10 & 0.10 \\
J04161210 & 0.20 & 0.19 & 0.19 & 0.19 & 0.19 & 0.14 & 0.12 \\
J04213459 & 0.12 & 0.35 & 0.18 & 0.17 & 0.14 & 0.10 & 0.10 \\
J04380083 & 0.05 & 0.50 & 0.50 & 0.50 & 0.50 & 0.50 & 0.30 \\
J04442713 & 0.05 & 0.18 & 0.17 & 0.16 & 0.15 & 0.12 & 0.10 \\
J04554535 & 0.20 & 0.26 & 0.26 & 0.26 & 0.26 & 0.16 & 0.10 \\
\enddata
\tablecomments{These values represent the 3$\sigma$ limits on
  q, the mass ratio,
as a function of distance from the primary for each of the
targets found to be single in our sample.}
\end{deluxetable}

The observed properties can be used to estimate the component
masses and mass ratios using pre-main sequence evolutionary
models.  Luhman (2004a) and Guieu et al. (2006) list for each
target an estimated extinction value A$_{V}$ (in the case of
the binaries, we assume this value is the same for both
components).  The absolute K-band magnitudes for the targets
were calculated using these values, m$_{K}$, and a distance of
140 $\pm$ 10 pc (Bertout et al. 1999) for Taurus.  The
unresolved spectral types were used to estimate the
temperatures of the brighter component of each pair, via the
temperature scale in Luhman et al. (2003).  Thus, using
M$_{K}$ and these temperatures, a mass and age for each
primary was found using the theoretical models of Baraffe et
al (1998).  Subsequently, the masses of the secondary
components were found via interpolating along the isochrone
derived for the primary to find the mass consistent with the
M$_{K}$ values for each secondary.  The masses derived from
this method are given in Table 2 and Table 3.  As shown in
Figure 2, all five secondaries are likely to be substellar, in
addition to one out of five of the primaries.

 For the majority of the targets, the calculated ages are
consistent with the age of Taurus.  However, a few targets
have an M$_{K}$ and temperature that predict an age younger
than 1 Myr (see Figure 2), which is beyond the range covered
by the Baraffe et al. (1998) models, but is occasionally seen
among T Tauri stars (Kenyon $\&$ Hartmann 1995).  Hence, in
the case of these systems, we fix the age to 1 Myr and use
this, with the temperature of the target, to estimate its
mass.  One of the binaries, CFHT-Tau 18, falls into this
category.  In this case, a new estimate of M$_{K}$ for the
primary component was derived by fixing its age, which was
then used to derive the mass of the secondary (via the
$\Delta$K calculated from the flux ratio).  Since the model
masses are quite independent of age and luminosity at this
stage of evolution, this age adjustment has negligible impact
on the derived masses.

\section{Discussion}

  The discovery of five binaries in a sample of twelve VLM
  objects increases the number of known, young, VLM binaries
  by $\thicksim$50$\%$.  This enables a more statistically
  robust assessment of VLM binary properties.  In order to
  derive an unbiased binary star fraction from our
  magnitude-limited survey (m$_{k}$ $\lesssim$11), we remove
  the one binary, CFHT-Tau 7, that has a primary with m$_k$
  $>$ 11 when resolved into its constituent components.  This
  leaves four binaries in a sample of eleven objects.  Only
  two other similar surveys of young, VLM objects exist in the
  literature\footnote{Recently, Bouy et al. (2006) presented a
  number of new binaries they found in Upper Scorpius.  These
  binaries are quite interesting in that many are quite wide
  and a number are VLM.  However, they
  do not include in their work the complete list or number of targets
  they observed.  Thus, for the purposes of this paper, we
  cannot include these binaries in our binary fraction analysis.}.  Kraus et
  al. (2005) surveyed 12 VLM objects in Upper Scorpius and
  found three binaries, and more recently, Kraus et al. (2006)
  targeted 22 VLM objects in Taurus and found two binaries.
  Figures 1 and 2 plot the properties of the binaries from
  these two surveys, along with those found in our study.
  Combining all three surveys (eight binaries out of 45
  targets) yields a binary fraction of 18 $\pm$ 4$\%$, with \textit{no}
  evidence of a difference between the two star-forming
  regions.

The VLM binary fraction in nearby star-forming regions appears
to be higher than that found in the field.  To make a direct
comparison, we define a binary fraction over the same
separation (4 to 100 AU) and mass (0.04 to 0.2 M$_{\sun}$)
range covered by the combined young, VLM sample.
Specifically, the field statistics are culled from the works
of Close et al. (2003), Bouy et al. (2003), and Gizis et
al. (2003), which survey objects of late M and early to mid-L
spectral types, with the surveys of mid-M type stars of
Siegler et al. (2005) and Reid and Gizis (1997, using only the
stars of spectral type M5-M9 in their sample) and the T dwarf
survey of Burgasser et al. (2003); all surveys included are
reported to be complete to at least q $\thicksim$ 0.5.  This
produces a sample of 39 binaries among 219 objects, which
results in a bias-corrected (see Burgasser et al. 2003 for
method) field VLM binary fraction of 8 $\pm$ 2$\%$.  This
value is a factor of two less and 2.2$\sigma$ lower than the young, VLM binary
fraction.

\begin{figure}
\epsscale{1.0}
\plotone{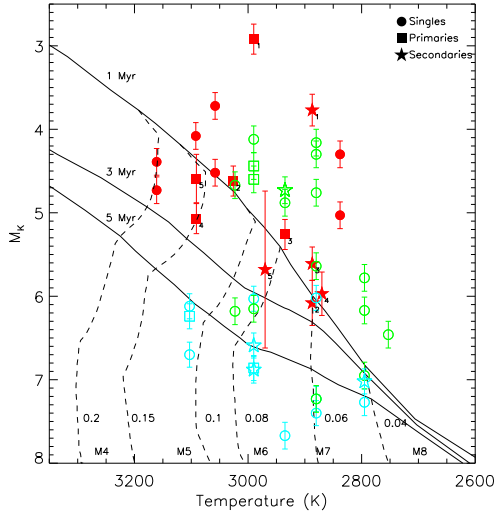}
\caption{The temperatures and absolute K-band magnitudes of
  each of our targets and those of the companions found
  (filled red symbols).  The primaries and secondaries that
  correspond to each other are numbered as such.  The
  isochrones shown are those of the age of Taurus,
  $\thicksim$1-5 Myr, and several mass tracks are also shown,
  as well as the spectral types that correspond to various
  temperatures (Luhman et al. 2003).  It is clear from this
  figure that several targets are substantially brighter than
  would be expected for members of Taurus, giving age
  estimates younger than 1 Myr (where the Baraffe et al. 1998
  tracks end). This lead us to use only the temperatures to
  estimate the appropriate mass, after fixing the age to
  1 Myr.  Also shown are those targets from other
  surveys of young, VLM stars that we use to construct an our
  entire sample of VLM targets from star-forming regions
  (green and blue open symbols). The green open symbols are
  sources also in Taurus, while the blue symbols are sources
  in Upper Scorpius, which appropriately cluster around
  slightly older isochrones than the Taurus members.}
\label{hrd}
\end{figure}

   In addition to the binary fractions discussed above, the
properties of VLM systems, namely the separation and the mass
ratio, show differences as a function of age.  Figure 1 plots
the binary mass ratio for the young and field surveys
discussed above, along with three other binaries discovered with
high angular resolution imaging (Luhman 2004b, Chauvin et
al. 2005, White et al. 1999, Bouy et al. 2006), as a function
of separation.  In contrast to the field population, young,
VLM binaries frequently have separations larger than 20 AU, as
well as smaller (more unequal) mass ratios.  This suggests
that the young VLM binaries are wider and have a flatter mass
ratio distribution than field binaries, or equivalently that
many of the young VLM binaries have smaller binding energies
than their older counterparts.

The differences in these parameters are quantified with both
one-dimensional and two-dimensional K-S tests, comparing the
separations and mass ratio distributions of these field VLM
binaries with the young VLM binaries.  In one dimension, the
separation distributions have a 1$\%$ probability of
similarity and the mass ratio distributions have a 0.02$\%$
probability of similarity.  Additionally, in two dimensions,
the distribution of both parameters taken together has only a
0.07$\%$ probability of similarity.  Thus, we can say with
a fairly high degree of certainty that the properties of young
VLM binaries differ substantially from those of old VLM
binaries.

These differences support the idea that
there may be substantial evolution in the properties of VLM
binaries $\thicksim$5-10 Myr after their formation (Burgasser
et al. 2006).  The disruption of binaries with separations
greater than 20 AU via interactions with their environment,
i.e. their formative cluster, shortly after their initial
formation would seem to be a plausible zeroth-order
explanation for the disparity.  As further evidence of this
idea, if we assume that evolution will eventually lead to the
disruption of the four binaries found here with separations
greater than 20 AU, we can calculate a new binary fraction -
the fraction that will survive to eventually become a field
binaries.  This leaves a total of four out of 45 binaries,
which yields a binary fraction of 9 $\pm$ 5$\%$.  This number
is perfectly consistent with the field binary fraction we
calculate above.
 
\begin{figure}
\epsscale{1.0}
\plotone{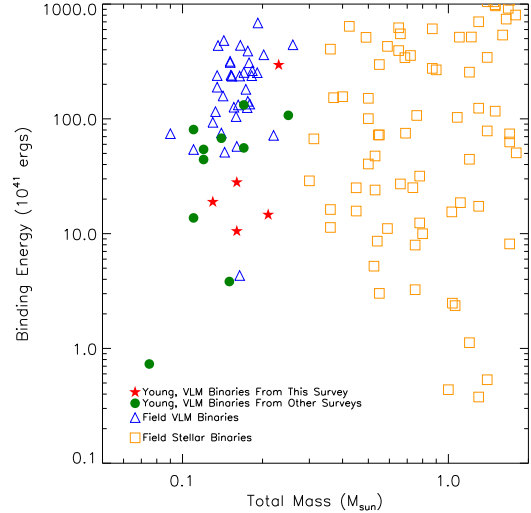}
\caption{Binding energy as a function of total mass for the
  VLM binaries, both young and in the field, discussed above,
  in addition to a number of known, stellar binaries (Close et
  al. 1990, Duquennoy $\&$ Mayor 1991, Fischer $\&$ Marcy
  1992, Reid $\&$ Gizis 1997, Reid et al. 2001).  Though three
  of the new binaries in Taurus have binding energies below
  the limit from Close et al. (2003), they do not have binding
  energies as low as those of stellar binaries.}
\label{be}
\end{figure}

However, the situation becomes substantially more complicated
when comparing VLM binaries to stellar binaries.  As noted by
both Close et al. (2003) and Burgasser et al. (2006), there
appears to be a discrepancy between the minimum observed
binding energy of stellar binaries and VLM binaries.  This
binding energy discrepancy is shown in Figure \ref{be}.  Three
of the binaries discovered here have a binding energy below
the previously determined limit, with a minimum around
$-$10$^{42}$ erg.  This minimum is still quite discrepant from
what is observed amongst field binaries, which appear to have
a minimum binding energy of about $-$10$^{41}$ erg.  The
stellar binding energy cutoff can be explained by the
work of Weinberg et al. (1987), who show that normal
interactions with the other stars and giant molecular clouds
in the Galaxy typically provide a ``velocity kick'' of less
than 1 km s$^{-1}$, sufficient to truncate a stellar mass
binaries with separations beyond $\thicksim$0.1 pc.  However,
generating a binding energy cutoff like that observed amongst
the VLM binaries necessitates a velocity kick roughly three
times this value.

One method of generating the higher velocity kicks for VLM
objects is from the
ejection scenario for VLM star and brown dwarf formation (Bate et
al. 2002), which predicts typical velocity kicks of
$\thicksim$3 km s$^{-1}$ very early in the formation process.
However, numerous authors have noted the implausibility of the
ejection scenario, as it predicts not only a very low
frequency of binary brown dwarf systems (5$\%$ or less), but
also generates binary brown dwarfs with separations of less
than 10 AU.  Although recent work by Umbreit et al. (2005)
shows that it is possible to form more brown dwarf binaries
via ejection than initially assumed, they find a semi-major
axis distribution that is severely truncated at wider
separations, with no brown dwarf binaries created at
separations greater than 20 AU.  While this new study initally
seemed a promising solution to the binding energy problem, it
does not explain the young, wide binaries plotted in Figure
\ref{young_compare}.

A reasonable explanation of the binding energy discrepancy
must now include the survival of some VLM binaries of
separations greater than 20 AU up through at least a few
million years after formation while at the same time account
for the truncation seen in the field.  On a number of
occassions, the low-mass, low-density nature of Taurus has
been invoked to explain difference observed between the
multiplicity properties of the region and regions like the
high-density Orion.  Thus, it could be argued that the
multiplicity of VLM objects found in this study is not
directly comparable to field VLM objects, which would most
likely have formed in higher mass star forming regions.
However, the similarity of the results presented here to those
found in Upper Scorpius make this an unsatisfactory
explanation - indeed, it has been argued that Orion is in fact
also an unusual star-forming region and most stars form in
intermediate mass regions like Upper Scorpius (e.g. Kroupa
1995).  Thus, the existence of wide VLM binaries in numerous
star-forming regions seems to be a significant trend.

The statistical significance of the trucation of wide, VLM
field binaries both now in terms of their younger counterparts
and stellar binaries therefore necessitates a plausible
physical explanation for their disappearance over a longer
timescale than has been previously suggested.  As mentioned
above, the work of Weinberg et al. (1987) suggests that
disruption of such binaries once they survive to become field
objects is unlikely.  In addition, although their eventual
disruption could potentially be due to interactions within
star forming regions before they disperse, Kroupa et
al. (2003) show from dynamical simulations that although many
VLM binaries would be disrupted in very high density regions
like the Orion Nebula cluster, the majority will not be
disrupted in regions like Taurus.  Disruption alone thus does
not appear to be the cause of the discrepancy.

However, the dynamical events a VLM binary undergoes
throughout its time in its formative cluster and
its eventual evolution into a field object is clearly not so
simple as it becoming either disrupted or not.  Most
interactions will have an effect on the properties of the
binary, namely in terms of hardening or softening.  Thus, it
could be that the impact of such interactions will be greater
on a VLM binary than on a stellar binary as a result of their
small size relative to the average perturbing object.  For
instance, Hills (1990) shows that the fractional change in
binding energy is a fairly strong function of the ratio of the
mass of the binary to the mass of the perturber, which would
imply that hard VLM binaries on average get ``harder'' more
quickly that stellar binaries.  Such parameter evolution,
first occuring in a star-forming region and then later in the
field, could potentially account for some of the binding
energy discrepancy.  For instance, the simulations of Adams et
al. (2006) show that interaction and disruption cross-sections
in a star-forming region scale roughly as the square root of
the mass of the primary star.  More detailed calculations that
take into account the effects of the relative mass of the
binary with respect to the rest of the cluster and
subsequently the field population should be performed to test
these effects.

\section{Summary}

   In a survey of thirteen newly-discovered VLM members of the
   Taurus star-forming region, we identified five new binary
   systems.  Follow-up observations are still required to
   confirm that these binaries are associated.  Proper motion
   measurements would provide the most definitive
   confirmation.  Additional constraints could be applied in
   the interim by obtaining either colors or spectral types
   for the secondaries in these systems.  Still, statistical
   arguments show that they are associated to a high degree of
   certainty.  As a result of these discoveries, we were able
   to statistically compare the properties of young, VLM
   binaries to their older counterparts in the field.  We
   found that our study and those of other young VLM objects
   suggest that the binary fraction of VLM objects is higher
   in star-forming regions like Taurus and Upper Scorpius than
   in the field.  Additionally, four of our five binaries have
   separations beyond 20 AU, a configuration previously found
   to be rare for VLM binaries.  These wide binaries, coupled
   with the statistically significant truncation of field VLM
   binaries beyond 20 AU, suggest that dynamical evolution
   produces the observed field VLM binary properties.  Along
   these lines, we find that it is difficult to produce these
   results using the ejection scenario of VLM object
   formation, as wide binaries are not expected to survive
   this process.  Dynamical perturbations may play a roll in
   determining the final distribution of VLM binary systems,
   as their small mass relative to the average mass of a
   perturbing object cam cause them to be more readily
   hardened over time than stellar mass binaries.  More
   simulations that take into the effects of the mass of a VLM
   binary with respect to the rest of either the cluster or
   the field population are needed to determine if such
   dynamics are sufficient to explain the binding energy
   discrepancy.

\acknowledgements

The authors thank observing assistants Steven Magee, Madeline
Reed, and Terry Stickel and support astronomer Mark Kassis for
their help in obtaining the observations and Elise Furlan and
Jessica Lu for their helpful suggestions.  We also thank an
anonymous referee for constructive feedback.  Support for this
work was provided by the NASA Astrobiology Institute and the
Packard Foundation.  QMK is supported by the NASA Graduate
Student Research Program (NNG05-GM05H) through JPL.  This
publication makes use of data products from the Two Micron All
Sky Survey, which is a joint project of the University of
Massachusetts and the Infrared Processing and Analysis
Center/California Institute of Technology, funded by the
National Aeronautics and Space Administration and the National
Science Foundation.  The W.M. Keck Observatory is operated as
a scientific partnership among the California Institute of
Technology, the University of California and the National
Aeronautics and Space Administration. The Observatory was made
possible by the generous financial support of the W.M. Keck
Foundation.  The authors also wish to recognize and
acknowledge the very significant cultural role and reverence
that the summit of Mauna Kea has always had within the
indigenous Hawaiian community.  We are most fortunate to have
the opportunity to conduct observations from this mountain.


\begin{thebibliography}{}
\bibitem[]{507} Adams, F.C., Proszkow, E.M., Fatuzzo, M., and
  Myers, P.C. 2006, \apj, 641,504
\bibitem[]{509} Ardila, D., Mart\'{i}n, E.L., \& Basri, G. 2000, \aj,
  120, 479
\bibitem[]{511} Baraffe, I., Charbier, G., Allard, F., \&
Hauschildt, P.H 1998, \aap, 337, 403 
\bibitem[]{513} Bate, M.R., Bonnell, I.A., Bromm, V., 2002, MNRAS,
332, 65
\bibitem[]{515} Bertout, C., Robichon, N., \& Arenou, F., 1999,
  \aap, 352, 574B
\bibitem[]{517} Bill\'{e}res, M., Delfosse, X., Beuzit, J-L.,
  Forveille, T, Marchal, L, \&  Mart\'{i}n, E.L. 2005, \aap,
  440, 55
\bibitem[]{520} Boss, A. 2002, \apj, 568, 743
\bibitem[]{521} Bouy, H., Brandner, W., Mart\'{i}n, E.L.,
  Delfosses, X., Allard, F., \& Basri, G, 2003, \aj, 126, 1526
\bibitem[]{523} Bouy, H.,  Mart\'{i}n, E.L., Brandner, W., \&
  Bouvier, J. 2005, \aj, 129, 511
\bibitem[]{525} Bouy, H., Mart\'{i}n, E.L., Brandner, W.,
  Zapatero-Osorio, M.R., B\'{e}jar, V.J.S., Schirmer, M.,
  Hu\'{e}lamo, N., \& Ghez, A.M. 2006, \aap, 451, 177
\bibitem[]{528} Brandner, W. et al., 2000, \aap, 120, 950
\bibitem[]{529} Burgasser, A.J., Kirkpatrick, J.D., Reid, I.N.,
  Brown, M.E., Miskey, C.L., Gizis, J.E., 2003, \apj, 586, 512
\bibitem[]{531} Burgasser, A.J., Reid, I.N., Siegler, N., Close,
  L., Allen, P., Lowrance, P., \& Gizis, J., 2006, Protostars
  and Planets V., ed. Reipurth et al., Univ. Arizona Press
\bibitem[]{534} Christou, J.C., 1991, PASP, 102, 1040
\bibitem[]{535} Close, L.M., Richer, H.B., \& Crabtree, D.R., 1990,
  \aj, 100, 1968
\bibitem[]{537} Close, L.M., Siegler, N., Freed, M., Biller, B.,
  2003, \apj, 587, 407
\bibitem[]{539} Duquennoy, A. \& Mayor, M., 1991, \aap, 248, 485
\bibitem[]{540} Fischer, D. A. \& Marcy, G.W., 1992, \apj, 396, 178
\bibitem[]{541} Fisher. R.T., 2004, \apj, 600, 769
\bibitem[]{542} Ghez, A.M., Neugebauer, G., \& Matthews, K. 1993,
  \aj, 106, 2005
\bibitem[]{544} Ghez, A.M., Weinberger, A.J., Neugebauer, G.,
Matthews, K., \& McCarthy, D.W., Jr. 1995, \aj, 110, 753
\bibitem[]{546} Gizis, J.E., Reid, I.N., Knapp, G.R., Liebert, J.,
  Kirkpatrick, J.D., Koerner, D.W., \& Burgasser, A.J., 2003,
  \aj, 125, 3302
\bibitem[]{549} Guieu, S., Dougados, C., Monin, J.-L., Magnier, E., \& Mart\'{i}n, E.L., 2006, \aap, 446, 485
\bibitem[]{550} Kenyon, S.J. \& Hartmanm, L. 1995, \apjs, 101, 117
\bibitem[]{551} Konopacky, Q.M., Ghez, A.M., McCabe, C., Duchene,
  G., \& Macintosh, B.A., 2007, \aj, in press
\bibitem[]{553} Kraus, A.L., White, R.J., \& Hillenbrand, L.A.,
  2005, \apj, 633, 452
\bibitem[]{555} Kraus, A.L., White, R.J., \& Hillenbrand, L.A.,
  2006, \apj, accepted
\bibitem[]{556} Kroupa, P., Bouvier, J., Duch\^{e}ne, G., \&
  Moraux, E. 2003, MNRAS, 346, 354
\bibitem[]{557} Labeyrie, A. 1970, \aap, 6, 85
\bibitem[]{558} Lu, J.R., et al., in prep
\bibitem[]{559} Luhman, K.L., 2004a, \apj, 617, 1216
\bibitem[]{560} Luhman, K.L., 2004b, \apj, 614, 398
\bibitem[]{561} Luhman, K. L., Stauffer, J.R., Muench, A.A.,
Rieke, G.H., Lada, E. A., Bouvier, J., \& Lada, C. J. 2003,
\apj, 593, 1093
\bibitem[]{564} Matthews, K., Ghez, A.M., Weinberger, A.J., \&
Neugebauer, G. 1996, \pasp, 108, 615
\bibitem[]{566} Matthews, K., \& Soifer, B. T. 1994, Infrared
Astronomy with Arrays, The Next Generation, ed. I. McLean 
(Dordrecht: Kluwer), 239
\bibitem[]{569}Mart\'{i}n, E.L., Barrado y Navascu\'{e}s, D.,
  Baraffe, I., Bouy, H., \& Dahm, S., 2003, \apj, 594, 525
\bibitem[]{571}Mart\'{i}n, E.L.,  Brandner, W., \& Bouy, H. 2006,
  \aap, submitted
\bibitem[]{573}Ne\"{u}hauser, R., Brandner, W., Alves, J.,
  Joergens, V., \& Comer\'{o}n, F., 2002, \aap, 384, 999
\bibitem[]{575} Padoan, P. \& Nordlund, A., 2004, \apj, 617, 559
\bibitem[]{576} Reid, I.N. \& Gizis, J.E. 1997, \aj,, 113, 2246
\bibitem[]{577} Reid, I.N., Gizis, J.E., Kirkpatrick, J.D., \&
  Koerner, D.W., \aj, 2001, 121, 489
\bibitem[]{579} Rice, W.K.M., Armitage, P.J., Bate, M.R., \&
  Bonnell, I.A. 2003, MNRAS, 339, 1025
\bibitem[]{581} Siegler, N., Close, L.M, Cruz, K.L., Mart\'{i}n,
  E.L, \& Reid, I.N. 2005, /apj, 621, 1023
\bibitem[]{583} Umbreit, S., Burkert, A., Henning, T., Mikkola, S.,
  \& Spurzem, R. 2005, \apj, 623, 940
\bibitem[]{585} Weinberg, M.D., Shapiro, S.L, \& Wasserman,
  I. 1987, \apj, 312, 367
\bibitem[]{587} White, R.J., Ghez, A.M., Reid, I.N., \& Schultz,
  G. 1999, \apj, 520, 811
\bibitem[]{589} Whitworth, A.P, \& Zinnecker, H., 2004, \aap, 427, 299
\bibitem[]{590} Whitworth, A.P. \& Stamatellos, 2006, astro-ph, 0610039
\bibitem[]{591} Zapatero Osorio, M.R., B\'{e}jar, V.J.S.,
  Mart\'{i}n, E.L., Rebolo, R., Barrado y Navascu\'{e}s, D.,
  \& Bailer-Jones, C.A.L., 2000, Science, 290, 103
\end{thebibliography}
\end{document}